\documentclass[aps,pra,twocolumn,superscriptaddress,floatfix]{revtex4-2}
\usepackage{amsmath}
\usepackage{amssymb}
\usepackage{graphicx}
\usepackage{color}
\usepackage{bm}
\usepackage[colorlinks=true,citecolor=blue]{hyperref}

\newcommand{\ket}[1]{\vert{#1}\rangle}
\newcommand{\ketbra}[2]{{\vert #1 \rangle\langle #2\vert}}

\newcommand{\braket}[2]{\langle{#1}|{#2}\rangle}
\newcommand* {\bra}[1]{\ensuremath{\langle {#1} |}}

\begin{document}
	\title{Perfect revivals of Rabi oscillations and hybrid Bell states in a trapped ion}
	
		\author{Juan Mauricio Torres}
	\email{jmtorres@ifuap.buap.mx}
	\affiliation{Instituto de F\'isica, Benem\'erita Universidad Aut\'onoma de Puebla, Apartado Postal J-48, Puebla 72570, Mexico}
	\author{Christian Ventura-Vel\'azquez}
	\affiliation{Instituto de F\'isica, Benem\'erita Universidad Aut\'onoma de Puebla, Apartado Postal J-48, Puebla 72570, Mexico}
	\author{Ivan Arellano-Melendez} 
	\affiliation{Facultad de Ciencias F\'isico Matem\'aticas, Benem\'erita Universidad Aut\'onoma de Puebla, Apartado Postal J-48, Puebla 72570, Mexico}
    	\affiliation{Instituto de F\'isica, Benem\'erita Universidad Aut\'onoma de Puebla, Apartado Postal J-48, Puebla 72570, Mexico}

	\date{\today}
\begin{abstract}
We show that perfect revivals of Rabi oscillations are possible, under certain conditions, in the population inversion of a trapped ion. Based on this property, we find that Schrödinger cat states of the atomic motion are naturally generated by the unitary dynamics. Using a pair of symmetric and antisymmetric Schrödinger cat states of the motion, together with an electronic excited or ground state, we find that the interaction leads to four orthonormal maximally entangled states of the two partitions, which we identify as Bell states. We also study a quadratic Kerr-type evolution that is possible for short interaction times.
\end{abstract}
\maketitle

\section{Introduction}

The collapse and revival of Rabi oscillations is a distinctive quantum phenomenon \cite{Eberly1980,Shore1993} that occurs in the Jaynes-Cumings model describing the interaction of a two-level atom with the quantized electromagnetic field \cite{Jaynes1963}. It was experimentally verified for the first time in the context of the one-atom maser \cite{Rempe1987,Meschede1985}. These oscillations are present, for instance, in the excited-state population of the two-level atom. They collapse and cease to exist for some time, to later reappear after the so-called revival time. As a characteristic feature, the revivals are wider in time, and an initial excited state is never fully restored \cite{Shore1993,Fleischhauer1993}. For large photon number, using a coherent state approximation one can obtain in a simple form the approximate revival time \cite{Gea-Banacloche1991a}, and also explain the factorization of atomic and photonic states in the collapse region \cite{Gea-Banacloche1990}. This feature can also be generalized to multiatom attractor states \cite{Jarvis2009a} when considering the Tavis-Cummings model \cite{Tavis1968b}. In this context, the phenomenon has been suggested to aid in quantum information protocols of atomic qubits assisted by multiphoton states \cite{Torres2016b,Bernad2016a}, but with some limitations due to an imperfect revival effect. However, it has been observed that
systems that present perfect revivals for two-qubit systems can lead to a complete Bell measurement \cite{Gonzalez-Gutierrez2018a} and entangling operations \cite{Gomez-Rosas2021}   
that can serve for more efficient entanglement purification schemes \cite{Torres2024}.
One of such systems is based on the Buck-Sukumar theoretical model \cite{Buck1981} which has not been implemented experimentally, while other proposals, more experimentally viable, rely on a two-photon process \cite{Gonzalez-Gutierrez2018a}, and trapped ions \cite{Gomez-Rosas2021}.

In trapped ions, collapses and revivals can also be present, as predicted and verified in Refs. \cite{Blockley1992,Cirac1994}, and  \cite{Meekhof1996}. In this case, the harmonic oscillator is represented by the quantized motion of the ion inside the trap, which couples to its electronic state through a laser field driving the atomic transition \cite{Leibfried2003a}. The fact that the situation in the Lamb-Dicke regime is well described by the Jaynes-Cummings model \cite{Cirac1994} not only explains the presence of this phenomenon, but it was also exploited in proposal \cite{Cirac1995}  and realization \cite{Schmidt-Kaler2003} of the Ciraz-Zoller gate
that illustrated the potential and positioned the ion trap architecture among the leading platforms for quantum computation \cite{Haeffner2008,Bruzewicz2019}. However, the Jaynes-Cummings model only describes a limited regime of trapped ions \cite{Leibfried2003a}. 
The system is more general and has led to several other applications for quantum information such as
heralded entanglement with ions in a cavity \cite{Casabone2013}, deterministic Bell states of ions \cite{Solano1999a}, and many others
\cite{Molmer1999,Roos2008,Blatt2008,Schmidt-Kaler2003}. Within the rotating wave approximation, the ion trap interaction model \cite{Vogel1995} results in a nonlinear Jaynes-Cummings model displaying much richer dynamics considered, for instance, to simulate the quantum Rabi model \cite{Pedernales2015}, the generation of large coherent states \cite{Alonso2016} and Schrödinger cat states \cite{Li1999}, and Kerr-type evolutions
\cite{Stobinska2011}.

In this work, we consider the non-linear Jaynes-Cummings model for a single trapped ion \cite{Vogel1995} and study the conditions in which perfect revivals of Rabi oscillations occur. 
We focus on large coherent states of the motion and find that the effect persists whenever the mean value of the coherent state is centered in an interval of quanta with a linear dependence of the eigenfrequencies with the phonon number. Previously in \cite{Gomez-Rosas2021} we exploited this effect for entangling non-unitary operations in two ions, while here we concentrate on a single ion and entangling unitary gates between its quantized motion and electronic states. 
We also show an alternative to generate Schrödinger cat states, and consider quadratic eigenenergy dependence in order to generate Kerr-type evolution. 

The paper is organized as follows. In Sec. \ref{Model}, we briefly present the model, find its formal solution, and present an approximation of the state vector in terms of operators acting on the vibrational mode. The deduction is kept general to describe other type of nonlinear models. We study the peculiarities of the eigenfrequencies of the ion trap model in Sec. \ref{sec:Frequencies}, where we present the conditions that lead to linear and quadratic dependence of the eigenfrequencies on the excitation number. In Sec. \ref{sec:Linear}, we focus on the linear regime, where we demonstrate the generation of Schrödinger cat of the motion and hybrid Bell states. In Sec. \ref{sec:Qudratic}, we focus on the quadratic regime of the eigenfrequencies and study the conditions that lead to a Kerr-type behavior of the eigenfrequencies. Finally we draw the coclusions in Sec. \ref{sec:Conclusions}.

\section{Nonlinear Jaynes-Cummings model}
\label{Model}
In this section we present the effective  Hamiltonian of a generalized version of the Jaynes-Cummings model with a nonlinear
intensity dependent coupling. 
We identify the constant of motion that allows exact solvability and present an approximate solution for initial coherent states of the bosonic mode.  Similar general models have already been
considered in the past and their exact solution is known \cite{Buck1981,Kochetov1987,Chaichian1990,Bonatsos1993,Rybin1999,SantosSanchez2016,Vogel1995}. 
However, here we are interested
in presenting a general approximate solution for initial coherent states with large mean number of quanta, which is especially convenient to analyze the entanglement in the system.

\subsection{Model and formal solution}
The ion-trap model \cite{Vogel1995} has the great virtue of being able to adapt to many situations. Under certain conditions, it can be reduced to the Jaynes-Cummings model, but it can also describe other situations involving more that one photon transition. In this work we will focus on the single photon transition, and in this section we will consider the following generic interaction Hamiltonian 
\begin{equation}
	\label{eq:V}
 V=\hbar \Omega
\left(
f(a^\dagger a)a\sigma_++a^\dagger f(a^\dagger a) \sigma_-
\right),
\end{equation}
where $a^\dagger$ and $a$ are the creation and annihilation
operators of the oscillator respectively. Furthermore, $f(a^\dagger a)$  is an intensity-dependent function with the number operator $a^\dagger a$. In this work, this mode corresponds to the quantized motion of the ion. In addition,
$\sigma_\pm$ are the raising and lowering operators of the ground states $\ket{g}$ and excited states $\ket{e}$ of the electronic degree of freedom of the ion.
 The Hamiltonian commutes with the constant of motion $I=a^\dagger a+\sigma_+\sigma_-$, having pairs of eigenstates $\{\ket{n-1,e},\ket{n,g}\}$ with eigenvalue $n>0$, and the singlet $\ket{0,g}$ for $n=0$.
 As $V$ commutes with $I$, the interaction Hamiltonian can be block-diagonalized in the aforementioned basis
 with the blocks given by
\begin{equation}
	\label{eq:BlockV}
    V^{(n)}=\hbar\left(
    \begin{array}{cc}
       0&\omega_{n}   \\
         \omega_{n}&   0
         \end{array}
    \right),\quad \omega_n= \Omega \bra{n}a^\dagger f(a^\dagger a)\ket{n-1}.
\end{equation}
%where $\omega_n$ is a real valued parameter dependent on $n$ and given by 
%\begin{equation}
%\label{eq:Omega}
%\omega_n= \Omega \bra{n}a^\dagger f(a^\dagger a)\ket{n-1}.
%\end{equation}
It follows from this block-diagonal structure that the system admits an exact solution. For $n>0$,  the eigenvalues and eigenvectors of $V$ are given by
\begin{equation}
    E_n^\pm=\pm\hbar\omega_n,\quad
    \ket{E_n^\pm}=\frac{\ket{e,n-1}\pm\ket{g,n}}{\sqrt{2}}.
\end{equation}
For $n=0$, the singlet $\ket{0,g}$ is an eigenstate with vanishing energy $E_0=0$. 

Having solved the eigenvalue problem, it is now possible to evaluate the time-dependent state vector for any given initial condition. We will
concentrate our attention on an initial product state
with the mode in the coherent state
\begin{equation}
	\label{eq:CoherentState}
 \ket{\alpha}=\sum_{n=0}^{\infty}p_{n}e^{in\phi }\ket{n}, \quad
	p_n=
	\frac{r^ne^{-\frac{|\alpha|^2}{2}}}{\sqrt{n!}},\quad\alpha=r e^{i\phi},
\end{equation}
where the complex parameter $\alpha$ has been expressed in polar form with its phase $\phi$  and the real parameter $r$. The electronic state is considered to be arbitrary as $\ket{\psi_0}$, and initial bipartite state is given as
\begin{equation}
\label{eq:InitialState}
	\ket{\tilde\Psi(0)}=\ket{\psi_0}\ket{\alpha}, \quad
    \ket{\psi_\phi}=c_g\ket{g}+
    c_e e^{-i\phi}\ket{e},
\end{equation}
where $c_e$ and $c_g$ denote the initial probability amplitudes of the excited and ground states respectively.
 In addition, we have also introduced the phase-dependent electronic state $\ket{\psi_\phi}$ in Eq. \eqref{eq:InitialState} as it allows us to rewrite the initial state as $\ket{\tilde\Psi(0)}=e^{i I\phi}\ket{\psi_\phi}\ket{r}$ in terms of a real coherent state $\ket{r}$.  Given the fact that $[I,V]=0$, one can find the following form of the time-dependent state
$\ket{\tilde\Psi(t)}=e^{-iVt/\hbar}\ket{\psi_0}\ket{\alpha}=e^{iI\phi}e^{-iVt/\hbar}\ket{\psi_\phi}\ket{r}$. This reduces the evaluation to real initial coherent states followed by a time-independent trivial transformation. In this way, the time evolution of the system is given by
\begin{align}
\label{eq:TimeStateExact}
    \ket{\tilde\Psi(t)}&=e^{iI\phi}
    \sum_{n,\pm}
    \frac{p_{n-1}c_ee^{-i\phi}\pm p_nc_g}{\sqrt{2}}
    e^{\mp i\omega_n t}\ket{E_n^\pm},
\end{align}
where $\sum_\pm$ denotes the summation over the plus and minus case. This form is exact with the convention that for $n=0$ there is only one state $\ket{E_0}$ leading to the single term $c_gp_0\ket{0,g}$. 
Taking $\phi=0$, the solution is also valid for any probability amplitudes $p_n$
that might be arbitrarily different from a coherent state. The form introduced here in terms of $\phi$ with real $p_n$ will prove useful in expressing the solution in terms of coherent states, as shown in what follows. 

\subsection{Approximation for large coherent states}
Let us now consider the case where the initial coherent state presents a large mean occupation number, i.e., $\bra{\alpha}a^\dagger a\ket \alpha= |\alpha|^2=r^2=N>>1$. In such a situation, one can use the following approximation of the probability amplitudes $p_{n-1}\approx p_n$, taking into account that we have chosen them real as in Eq. \eqref{eq:CoherentState}. Furthermore, one can take $p_0\approx0$ for large $N$. With these considerations, and by introducing the notation $\ket{l=0,1}$ to stand for $\ket{g}$ and $\ket{e}$ respectively, we can arrive from Eq. \eqref{eq:TimeStateExact} to the following expression
\begin{equation}
\label{eq:TimeStateApproxn}
   \ket{\tilde\Psi(t)} \approx 
   e^{iI\phi}\sum_{n,l}
    \frac{d_\phi^+e^{-i\omega_{n+l}t} -(-1)^l
    d_\phi^-e^{i\omega_{n+l}t}}{\sqrt{2}}p_n
    \ket{l}\ket{n}.
\end{equation}
We have also introduced the initial probability
amplitudes $d_\phi^\pm$ of the orthogonal states $\ket{\psi_\phi^\pm}$  that read
\begin{equation}
	\label{eq:phistates}
\ket{\psi_\phi^\pm}=\frac{e^{i\phi}\ket{e}\pm\ket{g}}{\sqrt{2}},\quad
	    d_\phi^\pm=\frac{e^{-i\phi}c_e
		\pm c_g
	}{\sqrt{2}}.
\end{equation}
Note that  the initial electronic state in Eq. \eqref{eq:InitialState} can be written as
$\ket{\psi_0}=d_\phi^+\ket{\psi_\phi^+}+d_\phi^-\ket{\psi_\phi^-}$, and therefore, at time $t=0$ one recovers the initial state in Eq. \eqref{eq:TimeStateApproxn}. In order to distinguish the approximate solution from its exact form, in the following we will employ $\ket{\Psi(t)}$ as the approximation to the exact state $\ket{\tilde\Psi(t)}$, where both states coincide at $t=0$.

By realizing that $\omega_{a^\dagger a}\ket n=\omega_n\ket n$, 
one can introduce the mode operators 
\begin{equation}
\label{eq:OperatorsA}
    A_l=\omega_{a^\dagger a+l},%-\omega_{a^\dagger a},\quad B=\omega_{a^\dagger a}
\end{equation}
that are diagonal in the number basis. Based on them, one is able to obtain the time-dependent state vector in the following form
\begin{equation}
	\label{eq:TimeStateOperators}
	\ket{\Psi(t)}=\frac{		\ket{g}\ket{\chi_0(t)}+e^{i\phi}\ket{e}\ket{\chi_1(t)}
	}{\sqrt{2\mathcal{N}(t)}},
\end{equation}
where $\mathcal{N}(t)$ is normalization constant that, which follows from the unnormalized mode time-dependent states
\begin{equation}
    \label{eq:ChiStatesOperators}
    \ket{\chi_l(t)}=d_\phi^+ e^{-i A_l t}\ket\alpha
    -(-1)^ld_\phi^- e^{iA_lt}\ket{\alpha}.
\end{equation}
One should remember that $\phi$ is the phase of the coherent state, namely $\alpha=re^{i\phi}$.  The deduction is general in the sense that we have not assumed any specific form of  $f(a^\dagger a)$ nor of the corresponding eigenfrequencies $\omega_n$
given in Eq. \eqref{eq:BlockV}. For instance, for the Jaynes-Cummings model one has that $f(a^\dagger a)=1$ and therefore $A_l=\Omega\sqrt{a^\dagger a+l}$. The states $\ket{\chi_l(t)}$ are, in general, not normalized, therefore one has to introduce the normalization constant 
\begin{equation}
\label{eq:Normaliztion}
\mathcal{N}(t)=
\frac{
\braket{\chi_0(t)}{\chi_0(t)}+\braket{\chi_1(t)}{\chi_1(t)}
}{2}=1+F_-(t),
\end{equation}
where we have considered the following pair of functions
\begin{equation}
\label{eq:Functions}
    F_\pm(t)={\rm Re}\left[
    d_\phi^{+\ast}d_\phi^-\bra\alpha
    (e^{i2A_1t}\pm e^{i2A_0t})
    \ket\alpha
    \right].
\end{equation}
These functions are also useful in evalauting the 
population inverstion $W(t)=\langle\sigma_z(t)\rangle$,
with $\sigma_z=\ketbra{e}{e}-\ketbra{g}{g}$, that 
takes the simple form
\begin{equation}
\label{eq:PopulationInvertion}
\hspace{-.1cm}
W(t)=
\frac{
	\braket{\chi_1(t)}{\chi_1(t)}-\braket{\chi_0(t)}{\chi_0(t)}
}{\braket{\chi_0(t)}{\chi_0(t)}+\braket{\chi_1(t)}{\chi_1(t)}}=
\frac{F_+(t)}{1+F_-(t)}.
\end{equation}

The result in Eq. \eqref{eq:TimeStateOperators} will be very useful in the following sections. However, there is another alternative  that makes more evident the possibility of obtaining a transient product state during the dynamics, namely
\begin{equation}
\label{eq:TimeStateOperators2}
    \ket{\Psi(t)}=\sum_\pm
    d_\phi^\pm
    \frac{\ket{e}e^{i(\phi \mp A_1 t)}\pm e^{\mp i A_0 t}\ket{g}}{\sqrt{2\mathcal{N}(t)}}
    \ket{\alpha}.
\end{equation}
This form of the solution will prove useful when we discuss the possibility of implementing a Kerr evolution of the mode in Sec. \ref{sec:Qudratic}.
\section{The ion trap model and its eigenfrequencies}
\label{sec:Frequencies}

\begin{figure}[!t]
\includegraphics[width=0.45\textwidth]{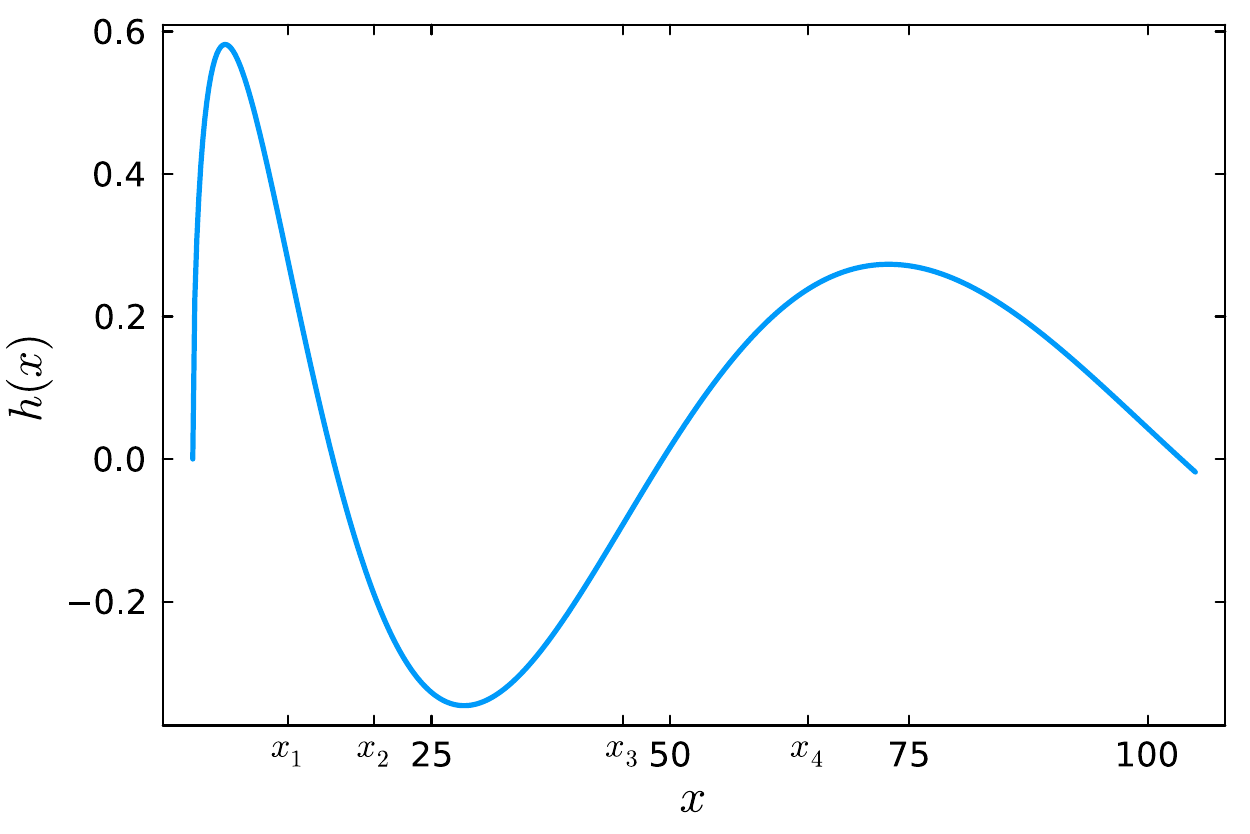}
	\caption{\label{fig:bessel} Bessel function of first kind $J_1(\sqrt{x})$ that determines the eigenfrequencies of the system $\omega_n$ in \eqref{eq:OmegaDefinitions}. A linear (quadratic) behavior in the eigenfrequencies is presented around the odd (even) values $x_j$ given in Table \ref{tab:numbers}.} 
\end{figure}

The intensity dependent function in the ion
trap model has the following form      written in normal order
\begin{equation}
\label{nTC:Hamiltonian}
f(a^{\dagger}a)=\eta e^{-\eta^2/2}\sum_{m=0}^{\infty}\frac{(-\eta^2)^m}{m!(m+1)!}a^{\dagger m}a^{m},
\end{equation}
where $\eta$ is the Lamb-Dicke parameter. 
With this expression, the eigenfrequencies in Eq. \eqref{eq:BlockV}
can be computed in terms of Laguerre polynomials
\begin{equation}
\label{eq:OmegaLagguerre}
	\omega_n=\frac{\Omega\eta e^{-\eta^2/2}}{\sqrt{n}} L_{n-1}^{(1)}(\eta^2)\simeq \Omega J_1(2\eta\sqrt{n}),\quad n\ge 1.
\end{equation}
The last expression is an approximation relying on the Bessel function of first order $J_1$ that is valid for small values of the Lamb-Dicke parameter $\eta$. In order to analyze the behavior of the eigenfrequencies, we introduce
the following definitions
\begin{equation}
\label{eq:OmegaDefinitions}
	\omega_{n}=\Omega h(x),\quad h(x)\equiv J_1(\sqrt x), \quad
	x=4\eta^2 n.
\end{equation}
Using them, one can focus on the behavior of $h(x)$ depending only on $x$ instead of the two variables $\eta$ and $n$. In Fig. \ref{fig:bessel} we plot  $h(x)$. In general, the behavior is highly non-linear, however, it can be noted that there is an apparent linear behavior around $x\approx10$. For this reason, we now consider a Taylor expansion around the mean phonon number $N=|\alpha|^2=r^2$ as
\begin{equation}
\label{eq:Taylor}
\omega_n=\omega_N+\omega_N'(n-N)+\frac{\omega_N''}{2}(n-N)^2+\dots
\end{equation}
We keep the second order term, we will also consider cases with quadratic dependence on $n$. Any  $k-$th order derivative of $\omega_n$ with respect to $n$ can be obtained from the derivatives of $h(x)$ as 
\begin{equation}
\label{eq:OmegaDer}
   \omega_n^{(k)}=\Omega(4\eta^2)^kh^{(k)}(x)
   =\Omega h^{(k)}(x)x^k/n^k.
\end{equation}
\begin{table}
\begin{tabular}{c|c|c|c|c}
$j$&1&2&3&4\\
\hline
	$x_j$&9.9516&19.014&45.068&64.469\\
     $h(x_j)$&0.279462&-0.19089&-0.091377&0.23870\\
     $h^{(1)}(x_j)$&-0.062845&-0.035116&0.022337&8.4062E-3\\
     $h^{(2)}(x_j)$&0&4.2246E-3&0&-1.0417E-3\\
     $h^{(3)}(x_j)$&1.3492E-3&0&-1.2091E-4&0\\
     $h^{(4)}(x_j)$&-3.6061E-4&-5.08E-5&7.9156E-6&3.9679E-6\\
\end{tabular}
\caption{\label{tab:numbers}
Four different values of $x_j$ and the derivatives of the function $h(x_j)$ determining the eigenfrequencies in Eq. \eqref{eq:OmegaDefinitions}.
The eigenfrequencies present approximately linear  (quadratic) behavior for odd (even) value of $j$ where $h^{(2)}(x_j)=0$ ($h^{(3)}=0$).  We have employed E-notation where ~-5.08E-5 $=-5.08\times 10^{-5}$.
}
\end{table}
The Taylor expansion is convergent, therefore, 
the coefficients of the expansion should be decreasing, and one can then expect certain separation of orders to different time scales in $\omega_n t$. For instance, one should expect a linear behavior in $n$, at least for short enough times, if one chooses a value of $N$ such that $\omega_N''=0$. In general, one can expect a $k-$th order behavior whenever
$\omega_N^{(k+1)}=0$. 
This $k$-th order will dominate in time then whenever the order $k+2$ remains negligible. This imposes the following restriction in time
\begin{equation}
\label{eq:TimeCondition}
t|\omega_N^{(k+2)}|\left(\sqrt{N}\right)^{k+2}/(k+2)!\ll 1,
\end{equation}
where we have replaced the term $n-N$ of the expansion by the standard deviation  of the dstribution, which in the present case of a Poisson distribution is given by $\sqrt{N}$.
The value of $\omega_N$ and its derivatives depends
on $N$ and the Lamb-Dicke parameter $\eta$.
The value of $N$ that fulfills $\omega_N^{(k+1)}=0$ can be obtained from Eq. \eqref{eq:OmegaDer} 
by determining the value of $x_j$ where $h^{k+1}(x_j)=0$. The connection between these two variables through $\eta$  follows from Eq. \eqref{eq:OmegaDefinitions} and is given by
\begin{equation}
    N=x_j/(4\eta^2).
\end{equation}
Here, the subindex $j$ labels different values.
In this work, we focus on linear and quadratic behavior, therefore in Table \ref{tab:numbers} we present two different zeros of the second (third) derivative with odd (even) $j$ leading to linear (quadratic) behavior.

\section{Linear regime}
\label{sec:Linear}

In this section, we study the dynamical features when a linear behavior dominates in the eigenfrequencies $\omega_n$, that is, when $\omega_N''=0$. In this case, the operator determining the evolution in Eq. \eqref{eq:TimeStateOperators} simplifies as
\begin{equation}
    \label{eq:OperatorLinear}
    A_l=\varphi_l+\omega_N' a^\dagger a,
    \quad
    \varphi_l=\omega_N+(l-N)\omega'_N.
\end{equation}
The photonic states in Eq. \eqref{eq:ChiStatesOperators} are now given in terms of
coherent states as
\begin{equation}
    \label{eq:ChiStatesLinear}
\ket{\chi_l(t)}=
    	d_\phi^+ e^{- i \varphi_lt}\ket{\alpha e^{-i \omega_N't}}-
	d_\phi^- e^{i (\varphi_l t+l\pi)}\ket{\alpha e^{i \omega_N't}}.
\end{equation}
We use the convention of revival time when the two components of the field coincide again in opposite side in phase space, namely $\ket{-\alpha}$. This happens
when $\omega_N't_r=\pi$ or simply
\begin{equation}
\label{eq:RevivalTime}
    t_r=\pi/|\omega_N'|=\pi N/|h'(x)|x\Omega,\quad
    t_h=t_r/2.
\end{equation}
We have also introduce the t
For the linear regime, we have chosen $\omega_N''=0$, therefore in the Taylor expansion, we require that the third order remains negligible. The restriction in Eq. \eqref{eq:TimeCondition} reduces to $|\omega'''_N\Delta_n^3|t/6 \ll1$, which translates into the following characteristic time 
\begin{equation}
\label{eq:TimeLinear}
	t\ll 
	\frac{6 |\omega_N'|t_r}{\pi|\omega_N'''\Delta_n^3| }
	%=\frac{6h'(x)x/N}{\pi h'''(x)x^3/N^3(2\sqrt{N})^3}
	\approx
	\frac{|h'(x)|\sqrt{N}}{4x_j^2|h'''(x)|}t_r.
\end{equation}
For $x_1$ this time is approximately $t_r\sqrt{N}/10$, and around four times smaller for $x_3$. For this reason, we will restric our attention to $x_1$, as the time constraint for $x_3$ is tighter and the required values for $N$ are larger.

\begin{figure}[!t]
\includegraphics[width=0.45\textwidth]{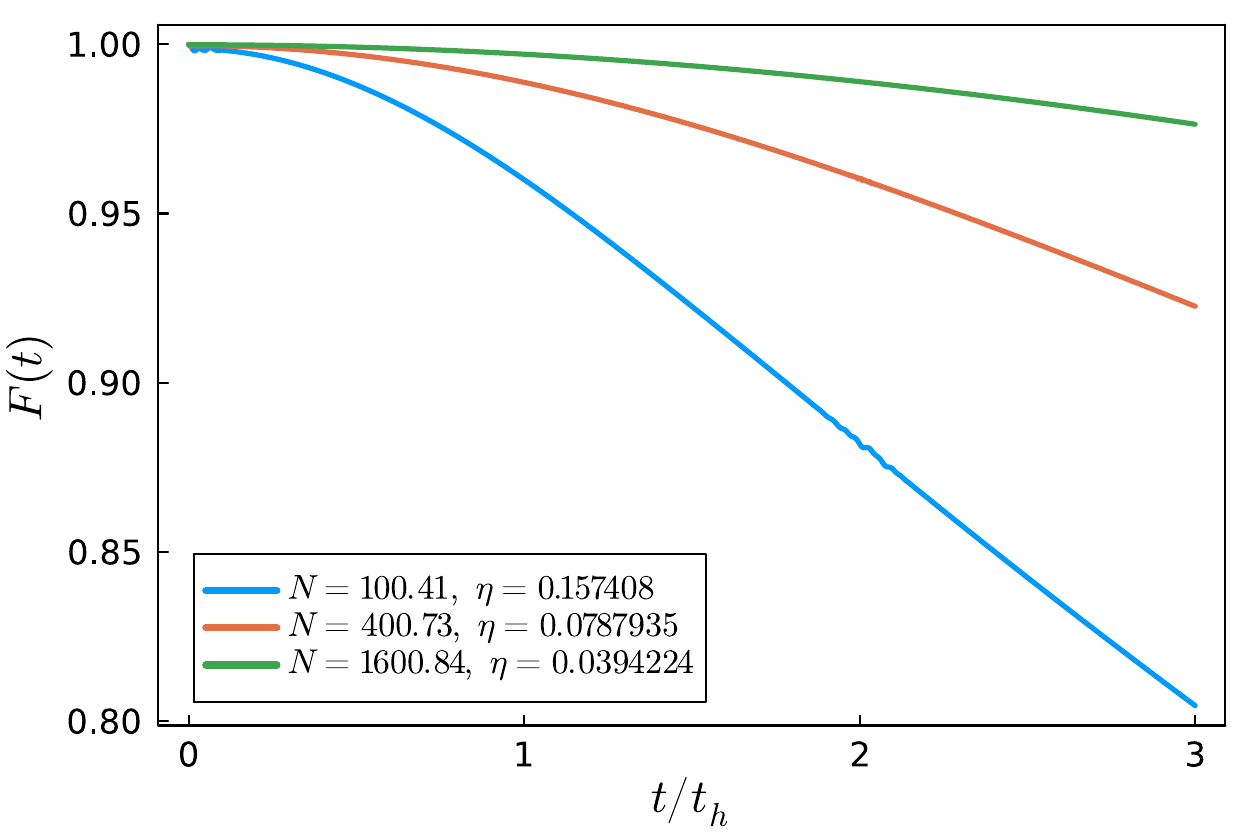}
	\caption{\label{fig:Fidelity} 
    Average fidelity $\overline{F(t)}$ between the exact and the analytical solutions in 
    Eq. \eqref{eq:TimeStateExact} and the approximate solution in Eq. \eqref{eq:TimeStateOperators} with the phononic states given for the linear regime given in Eq. \eqref{eq:ChiStatesLinear}. 
    Three different values of $N$ are shown and given in the inset. The average fidelity has been computed over one thousand random conditions keeping $N$ fixed.}
\end{figure}

In order to test our solution, we have evaluated the fidelity of the approximate $\ket{\Psi(t)}$ state with respect to the exact numerical calculation 
$\ket{\tilde \Psi(t)}$ in \eqref{eq:TimeStateExact} that can be evaluated as
\begin{equation}
\label{eq:Fidelity}
    F(t)=|\braket{\tilde\Psi(t)}{\Psi(t)}|^2.
\end{equation}
In Fig. \ref{fig:Fidelity} we have plotted the mean fidelity $\overline F(t)$ by taking the average fidelity over a thousand random initial condition, but with fixed $N$ as shown in the inset. The time has been rescaled in terms of half of the revival time $t_h$, so that $t=2t_h$ corresponds to the revival time. The fidelity increases with $N$ confirming the prediction in \eqref{eq:TimeLinear}.

\begin{figure}[!t]
\includegraphics[width=0.45\textwidth]{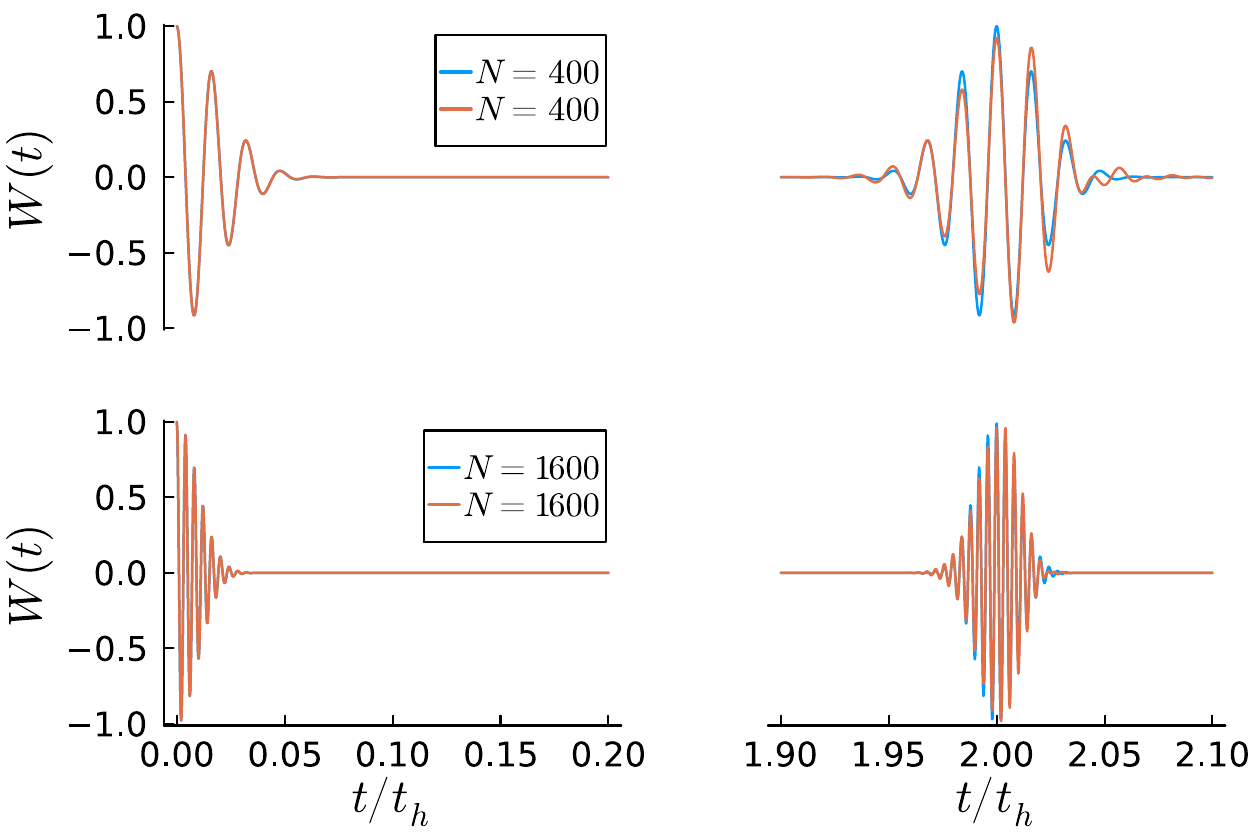}
	\caption{\label{fig:W}
    Population inversion $W(t)=\langle\sigma_z(t)\rangle$ presenting collapse and nearly perfect revivals of Rabi oscillations
    at the revival time $t_r=2t_h$ for two different values of the mean phonon number $N$.
    }
\end{figure}

The population inversion can be evaluated using the expressions in Eq. \eqref{eq:PopulationInvertion} and the functions for this linear case simplify as
\begin{equation}
\label{eq:FsLinear}
    F_\pm(t)={\rm Re}
    [d_\phi^{+\ast}d_\phi^{-}e^{i2\varphi_0 t}(
    e^{i2\omega_N't}\pm 1)
    e^{|\alpha|^2(e^{i2\omega_N't}-1)}].
\end{equation}
The result can be further simplified by realizing that 
$\omega_N/\omega_N'=N h(x_1)/x_1|h'(x_1)|>>1$ for lage $N$. In addition, one has 
that $F_-(t)\approx0$, and if one chooses an initial excited state $\ket e$, where $d_\phi^\pm=1/\sqrt2$, one can arrive at the following approximate expression for the population inversion
\begin{equation}
\label{eq:WLinearApp}
    W(t)=e^{N(\cos2\omega_N't-1)}
    \cos\left[2\varphi_1t+N\sin2\omega_N't\right].
\end{equation}
This expression is plotted in Fig. \ref{fig:W} in blue, together with the exact numerical calculation in orange, and for two different values of $N$. It should be noted that a better agreement is found for larger values of $N$ as expected from the previous discussion on the validity of the linear expansion. 

\subsection{Schrödinger cat states at half revival time}

An interesting behavior is found for interaction times equal to half the 
revival time $t_h$, where $e^{\pm i \omega_N't_h}=\pm i$. In this case, both 
 mode components, up to a phase,  coincide in the same state, namely $\ket{\chi_1(t_h)}=-i\ket{\chi_0(t_h)}$ with
\begin{equation}
\label{eq:chihalftr}
\ket{\chi_0(t_h)}=d_\phi^+ e^{-i\varphi_0t_h}\ket{-i\alpha }-
d_\phi^- e^{i\varphi_0t_h}\ket{i\alpha }.
\end{equation}
In this way, the bipartite state factorizes as a the product state
\begin{equation}
	\label{eq:Psihalftr}
\ket{\Psi(t_h)}=\frac{\ket{g}- i e^{i\phi}
	\ket{e}}{\sqrt{2}}
\ket{\chi_0(t_h)},
\end{equation}
where one can note that at this time $\mathcal{N}(t_h)\simeq 1$, as $F_\pm(t_h)\propto e^{-2|\alpha|^2}$ is vanishingly small.

The solution in Eq. \eqref{eq:Psihalftr} is valid at $t_h$ for an arbitrary electronic state and the vibration mode in any coherent state $\ket{\alpha}$. The superposition in the mode state can lead to a Schrödinger cat state if both probability amplitudes $d_\phi^\pm$ coincide in absolute value. There are also relative phases that
can be adjusted with the initial conditions of the electronic degrees of freedom. For example, one can get rid of these phases by
preparing the state in such a way that 
$d_\phi^\pm =e^{\pm i\varphi_0t_h}/\sqrt{2}$ or 
$d_\phi^\pm =\pm e^{\pm i\varphi_0t_h}/\sqrt{2}$. These two possibilities 
are contained  in the following global initial condition 
\begin{equation}
	\ket{\Psi^\pm(0)}=
	\frac{e^{i\varphi_0t_h}\ket{\psi_\phi^+}\mp e^{-i\varphi_0t_h}\ket{\psi_\phi^-}}{\sqrt 2}
	\ket{\alpha}.
\end{equation}
We remind the reader that $d_\phi^\pm$ are the initial probability amplitudes of the electronic states $\ket{\psi_\phi^\pm}$ given 
in Eq. \eqref{eq:phistates}.
After an interaction lasting half the revival time with this initial condition,
the  bipartite state results simply in
\begin{equation}
	\ket{\Psi^\pm(t_h)}=
	\frac{\ket{g}-ie^{i\phi}\ket{e}}{\sqrt{2}}
	\frac{\ket{-i\alpha }\pm \ket{i\alpha}}{\sqrt{2}}.
\end{equation}
With this procedure, and by adjusting the phase $\phi$, one is able to generate any Schrödinger cat state of the motional mode. 
Note that in this scheme, the cat states are generated solely by the coherent dynamics. There is no need to measure the electronic state \cite{Li1999}, nor to use a Raman sequence  \cite{Monroe1996} as in other proposals. 
%In this unitary dynamics leads to a bipartite product state.

Having explained the generation of cat states in the vibrational mode, we can now consider them as initial conditions forming a product state with an arbitrary electronic condition, that is
\begin{equation}
	\label{eq:initialEnt}
	\ket{\Psi_\pm(0)}=\ket{\psi_0}\frac{\ket{\alpha}\pm\ket{-\alpha}}{\sqrt{2}}.
\end{equation}
Using Eq. \eqref{eq:Psihalftr}, we can evaluate the time evolved state after a time $t_h$, as it suffices to employ the normalized superposition of the solutions for coherent states with phases $\phi$ and $\phi+\pi$. Furthermore, by noting that the electronic probability amplitudes fulfill the relation $d_{\phi+\pi}^\pm=-d_\phi^\mp$, one is able to arrive to the state 
\begin{align}
	\label{eq:Psipmtr2}
	\ket{\Psi_\pm(t_h)}=
	&\frac{e^{-i\varphi_0t_h}\pm e^{i\varphi_0t_h}}{\sqrt{2}}\ket{g}
	\frac{
    d_\phi^+\ket{-i\alpha}\mp d_\phi^-\ket{i\alpha}
    }{\sqrt{2}}
	\nonumber\\
	-ie^{i\phi}
	&\frac{e^{-i\varphi_0t_h}\mp e^{i\varphi_0t_h}}{\sqrt{2}}\ket{e}
	\frac{
    d_\phi^+\ket{-i\alpha}\pm d_\phi^-\ket{i\alpha}
    }{\sqrt{2}}
    .
\end{align}
The resulting bipartite state is, in general, not a product state anymore. In fact, it can be maximaly entangled if the amplitudes $|d_\phi^\pm|=1/\sqrt2$, and if $\varphi_0t_h=\pi M+\pi/4$ with $M\in \mathbb{Z}$, a condition that can be accopmlished by considering the eigenfrequencies in Eq. \eqref{eq:OmegaDefinitions} and the definition of $t_h$ in \eqref{eq:RevivalTime}. Doing so, one finds
\begin{equation}
    \label{eq:MagicN}
    N=\frac{x_j|h'(x_j)|}{h(x_j)-x_jh'(x_j)}(2M+1/2).
\end{equation}
Note that this result is independent of the Lamb-Dicke parameter $\eta$.
The sign of the integer number $M$ matches the sign in the denominator of \eqref{eq:MagicN}, since $N$ must be a positive number.
With this result, we will next show that it is possible
to obtain four maximally entangled orthogonal states, starting from four separable orthogonal initial conditions of the hybrid system.

\begin{figure}
\includegraphics[width=0.45\textwidth]{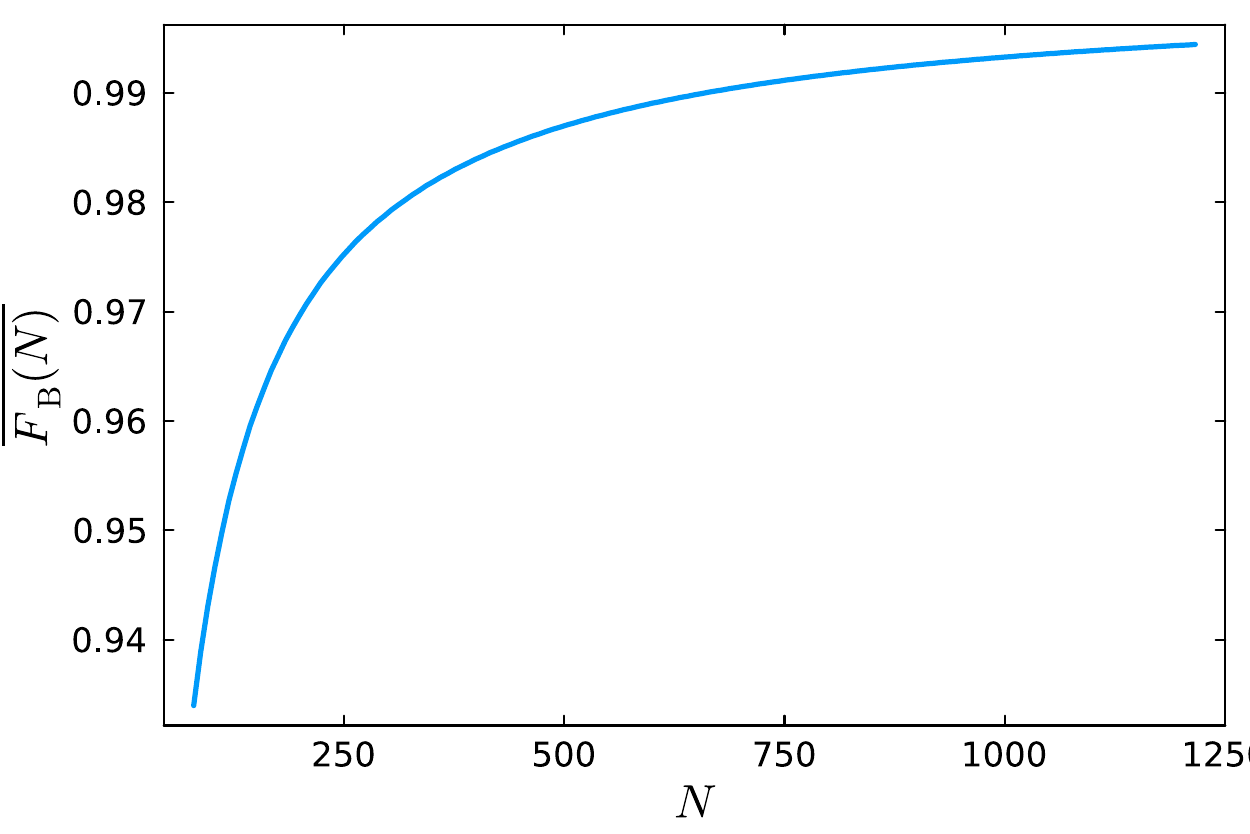}
	\caption{\label{fig:FidelityBell}
    Average fidelity of the hybrid Bell states generated by the exact time-evolution $\ket{\tilde\Psi(t)}$ with respect to the Bell states in  \eqref{eq:BellStates}  predicted by the approximation as a function of the mean phonon number $N$. The fidelity 
    has been averaged over $50$ different phases $\phi$ of the coherent state and among
    the four Bell states.
    }
\end{figure}

\subsection{Bell basis at half revival time}
Let us now consider four specific initial conditions:
the electronic states in either ground or excited state, and 
the mode in one of the following symmetric ($k=0$) or antisymmetric ($k=1$) Schrödinger cat states
\begin{equation}
\label{eq:CatStates}
	\ket{\alpha,k}=
	\frac{
		\ket{{\alpha}}
		+(-1)^k
		\ket{-\alpha}}{\sqrt{2}}.
\end{equation}
All possible combinations can be compactly written as
\begin{equation}
	\label{eq:initialBell}
	\ket{\Psi_l^k(0)}=\ket{l}\ket{\alpha,k}
	%\frac{\ket{\alpha}\pm\ket{-\alpha}}{\sqrt{2}}
	,\quad \ket{0}\equiv\ket{g},\quad \ket{1}\equiv\ket{e},
\end{equation}
where we have relabeled for convenience the electronic sates. Furthermore, we will assume  $\varphi_0t_h=\pi M+\pi/4$ with  integer $M$ as explained before.  Using these considerations and the result in Eq. \eqref{eq:Psipmtr2}, the 
final state can be calculated to arrive at 
\begin{equation}
\label{eq:BellStates}
	\ket{\Psi_l^k(t_h)}=
	\frac{
	\ket{g}\ket{i\alpha,l\oplus k}
+e^{i\pi k}e^{i\phi}
\ket{e}
\ket{i\alpha,l\oplus k\oplus 1}	
	}{\sqrt{2}}
\end{equation}
where $\oplus$ denotes the sum modulo $2$ and where we have omitted the overall phase $i^k(-1)^{l+M}e^{-i\phi}$. It can be verified that the states are orthonormal and maximally entangled. If the symmetric and antisymmetric cat states, $\ket{i\alpha,0}$
and $\ket{i\alpha,1}$, are considered as qubits, together with a second qubit encoded in $\ket e$ and $\ket g$, the four states in Eq. \eqref{eq:BellStates} can be regarded as a Bell basis. In Table \ref{BellTable}, we explicitly present all possible combinations of input and output states when $M$ is an even number.

In order to check the validity of the Bell states, we have numerically calculated in Fig. \ref{fig:FidelityBell} the average fidelity between the ideal states and the ones obtained from the exact dynamics as a function of $N$, namely
\begin{equation}
    \label{eq:FidelityB}
    F_B(N)=\sum_{k,l=0}^1|
    \bra{\Psi_l^k(t_h)}e^{-i Vt_h/\hbar}\ket{\Psi_l^k(0)}|^2.
\end{equation}
In Fig. \ref{fig:FidelityBell} we have in addition performed an additional average over $50$ different phases of the coherent states to obtain $\overline{F}_B(N)$. The fidelity values are larger than $0.93$ for all the values of $N$ starting from $80$. Even for low values of $N$ we achieve good fidelity. As the number of $N$ increases, the value asymptotically approaches unity. In particular, fidelity larger than $0.99$ is achieved for $N>750$.
\begin{center}
	\begin{table}
	\begin{tabular}{c|ccc|c}
	Input state & &Output state&&phase\\
	\hline
    	$\ket{g}\ket{\alpha,0}$ & &$\ket{g}\ket{i\alpha,0}+e^{i\phi}\ket{e}\ket{i\alpha,1}$&&$1$\\
	$\ket{g}\ket{\alpha,1}$ & 
	&$\ket{g}\ket{i\alpha,1}-e^{i\phi}\ket{e}\ket{i\alpha,0}$&&$i$\\
	$\ket{e}\ket{\alpha,0}$ & &$\ket{g}\ket{i\alpha,1}+e^{i\phi}\ket{e}\ket{i\alpha,0}$&&$-e^{-i\phi}$\\
	$\ket{e}\ket{\alpha,1}$ & &$\ket{g}\ket{i\alpha,0}-e^{i\phi}\ket{e}\ket{i\alpha,1}$&&$-ie^{-i\phi}$
	\end{tabular}
\caption{\label{BellTable}
Hybrid output Bell states generated by the dynamics of the system with its corresponding input separable state. The overall phase is shown in the third column. We have omitted the normalization factor $1/\sqrt2$ for a better readability. 
}
	\end{table}
\end{center}

\subsection{Phase space of the motional mode}
\begin{figure}
\includegraphics[width=0.45\textwidth]{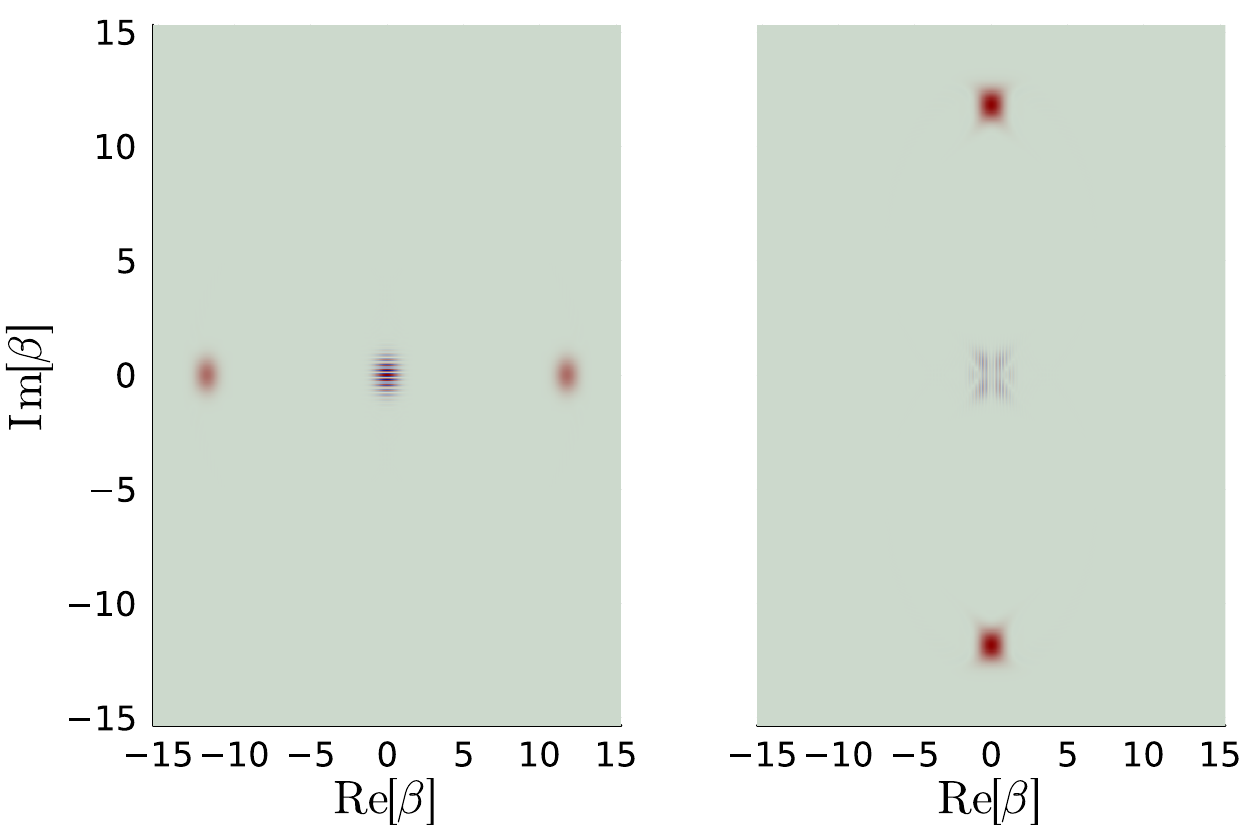}
	\caption{\label{fig:WignerBell}
    Left: Wigner function of the symmetric cat state $\ket{\alpha,0}$ in Eq. \eqref{eq:CatStates} that is the reduced phononic state of $\ket{\Psi_0^0(0)}$ in \eqref{eq:initialBell}. Right: Wigner state of the of the reduced density matrix at $t_h$ from the initial state $\ket{\Psi_0^0(0)}$. The interference fringes are very small showing the incoherent superposition of two coherent states.
    }
\end{figure}

To finalize this section, we focus on the visualization of the motional mode in phase space. We choose as pseudo-probability distribution, the Wigner function that takes the following form
\begin{equation}
    \mathcal{W}(\beta)={\rm Tr}\left\{2
    D(\beta)e^{i\pi a^\dagger a}D(-\beta)
\rho_{\rm ph}
    \right\},
\end{equation}
with the displacement operator $D(\beta)=\exp(\beta a^\dagger -\beta^\ast a)$, and
where $\rho_{\rm ph}={\rm Tr}_{\rm el}
\left\{\ketbra{\tilde\Psi(t)}{\tilde\Psi(t)}\right\}$ is the reduced density matrix of the motional partition after taking partial trace over the electronic degrees of freedom. In Fig. \ref{fig:WignerBell} we plot the Wigner function of a symmetric cat state $\ket{\alpha,0}$, with $\alpha=11.77$ in the left panel. The right plot shows the Wigner function of the reduced motional density operator from the state $\ket{\Psi_0^0(t_h)}$ corresponding to the first Bell state in the table \ref{BellTable}. Small interference fringes are still noticeable, given the fact that the mean number of phonons is not too large, $N=184.42$. The vanishing interference fringes corroborate the incoherent superposition of symmetric and antisymmetric cat states, which is equal to the incoherent superposition of two coherent states, namely $\ketbra{i\alpha}{i\alpha}+\ketbra{-i\alpha}{-i\alpha}$.

\section{Quadratic regime}
\label{sec:Qudratic}
In this section, we consider the possibility of achieving quadratic dependence of the eigenfrequencies in the excitation number $n$.
In order to fulfill this property, we will consider $\omega_N'''=0$, as in such a case, the Taylor expansion in Eq. \eqref{eq:Taylor} will not present a third order term in $n-N$, and for short enough times, when $\omega_N^{(4)}\Delta_n^4t/4!\ll 1$, the quartic term can be neglected. In this case, the operators in Eq. \eqref{eq:OperatorsA} are given by
\begin{align}
    \label{eq:OperatosAQuadratic}
    A_l&= 
    %\omega_N+\omega_N'(a^\dagger a+l-N)+\frac{\omega_N''}{2}(a^\dagger a+l-N)^2
   % \nonumber\\
    \delta_l+\Delta_la^\dagger a+\frac{\omega''_N}{2}\left(a^\dagger a\right)^2,\\
    \delta_l&=\varphi_l+\frac{\omega_N''}{2}(l-N)^2,
    \quad \Delta_l=\omega_N'+(l-N)\omega_N''
    \nonumber.
\end{align}
It is well known that a coherent 
state under quadratic time evolution, with a Hamiltonian proportional to $(a^\dagger a)^2$, does not maintain its coherent state form \cite{Stobinska2011}. However, there is a revival time where it returns to a single coherent state, namely 
$e^{-i (a^\dagger a)^2 \pi}\ket\alpha=\ket{-\alpha}$. Here, this time is achieved when 
\begin{equation}
\label{eq:t2}
	t_2=2\pi/|\omega_N''|=\frac{2\pi}{\Omega |h''(x_j)|(4\eta^2)^2}=
	\frac{2\pi N^2}{\Omega |h''(x)|x^2}.
\end{equation}
From Eq. \eqref{eq:TimeCondition}, it is required that $\omega_N^{(4)}\Delta_n^4t/4!\ll 1$ which implies that the interaction time has to fulfill the following condition
\begin{equation}
t\ll \frac{4!}{|\omega^{(4)}_N|\Delta_N^4}=
\frac{|\omega''_N| t_2 4!}{2\pi|\omega^{(4)}_N|\Delta_N^4}
%=\frac{12 h^{(2)}(x)x_j^2N^4 t_2}{2\pi h^{(4)}x_j^4N^2(x)(2\sqrt{N})^4}
=
\frac{3 |h^{(2)}(x_j)|}{8\pi |h^{(4)}(x_j)|x_j^2}t_2.
\end{equation}
Given in terms of $t_2$,
this condition is independent of the Lamb-Dicke parameter $\eta$ and the mean excitation number $N$. Therefore,  the second order approximation is, in general, only valid for times much shorter than the revival time $t_2$. For $x_2$ in table \ref{tab:numbers}, this translates into $t\ll  0.003 t_2$, meaning that the interaction time cannot be  close to $t_2$, independently of the values of $N$ and $\eta$. For $x_4$ this time is even shorter. For this reason, in the following we will focus our attention to short times where the approximation is valid.

\begin{figure}[!t]
\includegraphics[width=0.49\textwidth]{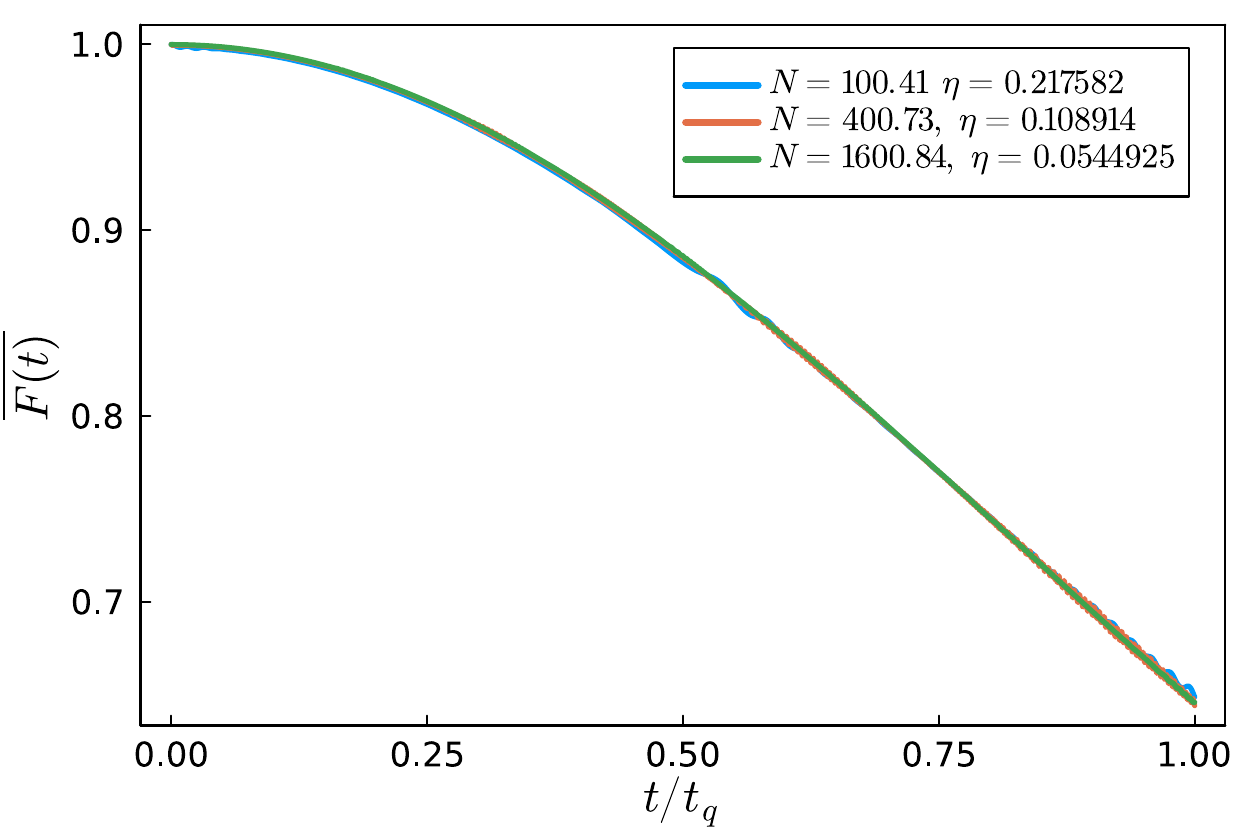}
	\caption{\label{fig:FidelityQuad}
      Average fidelity $\overline{F(t)}$ between the exact and the analytical solutions in 
    Eq. \eqref{eq:TimeStateExact} and the approximate solution in Eq. \eqref{eq:TimeStateQuadratic}. 
    Three different values of $N$ are shown and given in the inset. The average fidelity has been computed over one thousand random conditions keeping $N$ fixed. The three curves coincide, as we have escaled the time with $t_q$ in \eqref{eq:tq} determining the validity of the approximation.
    }
\end{figure}

\subsection{Short-time behavior}
Achieving long interacting times, such as $t_2$ might be challenging in experiments and are without the validity of our approximation. However, one might focus on small times where quadratic behavior is relevant, and even more on times where one is able to split quadratic Hamiltonians between different electronic attractor states \cite{Jarvis2009a}. To achieve this, we restrict our attention to the times where the operators $A_l$ in Eq. \eqref{eq:TimeStateOperators2} differ by a constant. In particular one finds that
\begin{align}
\label{eq:OperatorsQuadDif}
    A_1-A_0&=\omega_N'+\omega_N''(a^\dagger a-N+1/2)\approx\omega_N'.
\end{align}
The last term is an approximation only valid when
$|\omega_N''|\sqrt{N}t\ll1$, i.e., for times $t<<t_2/2\pi\sqrt{N}$. For convenience, we define
the characteristic time
\begin{equation}
\label{eq:tq}
    t_q=1/\sqrt{N}|\omega_N''|,
\end{equation}
 where the approximation in Eq. \eqref{eq:OperatorsQuadDif} is valid. 
Taking this into account in the compact form of the solution given in Eq. \eqref{eq:TimeStateOperators2},
one can write the state as
\begin{equation}
\label{eq:TimeStateQuadratic}
    \ket{\Psi(t)}=\sum_\pm
    d_\phi^\pm
    \frac{\ket{e}e^{i(\phi \mp \omega_N' t)}\pm \ket{g}}{\sqrt{2\mathcal{N}(t)}}
    e^{\mp iA_0 t}\ket{\alpha }.
\end{equation}

The relevant feature of this state is that two motion states evolve separately, accompanying two electronic states. For example, when choosing $d_\phi^+=1$, that is, $\ket{\psi_\phi^+}$ as an initial condition, the bipartite evolves as the state of the product
$\ket{\psi_{\phi-\omega_N't}^+}e^{- i A_0 t}\ket{\alpha}$. This opens the possibility of implementing a quadratic Kerr-type evolution to the motion state that can be relevant for quantum computation with continuous variable systems \cite{Lloyd1999}.

To test our approximate solution, in Fig. \ref{fig:FidelityQuad}, we plot the fidelity of the state in Eq. \eqref{eq:TimeStateQuadratic}
with respect to the exact numerical calculation. 
We have performed an average over a thousand initial conditions for three values of $N$ whose values is shown in the legends.
The time has been rescaled in terms of the characteristic time $t_q$ introduced in Eq. \eqref{eq:tq}. The three curves for different values of $N$ coincide, confirming the prediction given for $t_q$.

\begin{figure}[!t]
\includegraphics[width=0.49\textwidth]{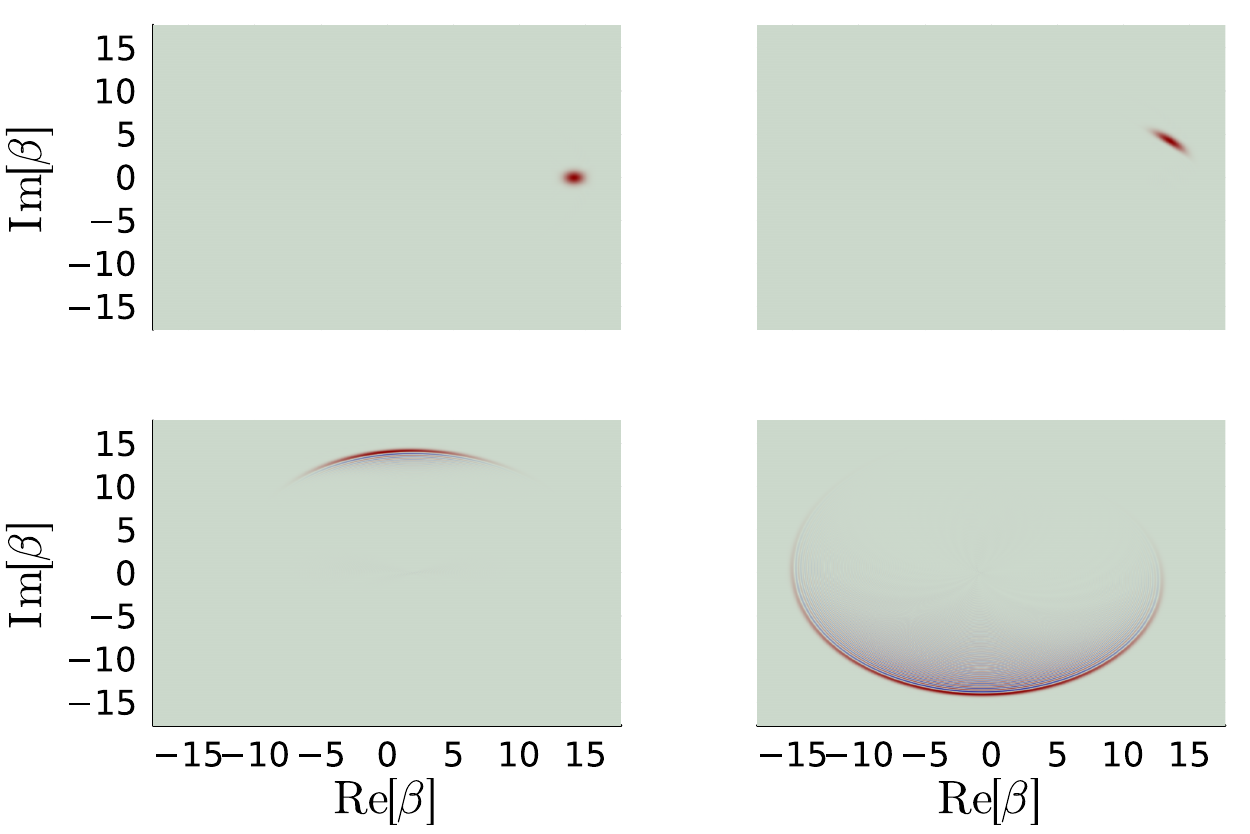}
	\caption{\label{fig:WignerQuad.pdf}
    Wigner function of the mode reduced density matrix from the state in \eqref{eq:TimeStateQuadratic} with $d_\phi^+=1$. Three different values of time where condsidered: $t=,\,t_q/4,t_q/2,\,3t_q/4$. $N=200$.
    }
\end{figure}

To verify the quadractic behavior of the motion states, in Fig. \ref{fig:WignerQuad.pdf}, we present the Wigner function of an initial coherent state and its time evolution for three different interaction times. One can note the deformation of the initial coherent state up to a point where it starts to cover a complete circle in phase space.

\section{Conclusions}
\label{sec:Conclusions}
We have presented an approximate solution for initial coherent states and general electronic states suitable for nonlinear Jaynes-Cummings models in terms of operators acting on the harmonic oscillator. For the ion trap model \cite{Vogel1995}, we find the conditions that admit a linear dependence of the eigenfrequencies on the excitation number and show that Schrödinger cat states of the motion are possible to achieve with the unitary dynamics. More importantly, we demonstrate the possibility to maximally entangle the quantized motion with the electronic degree of freedom in four orthogonal hybrid Bell states formed by the electronic states and a pair of symmmetric and antisymmetric Schrödinger cat states. 
We have also shown the conditions that lead to a separate Kerr-type evolution of the motion that could be useful in the context of quantum computation with continuous variable systems \cite{Lloyd1999}. 

All of this is possible, provided that the motion state is prepared in large coherent states such as in current experimental ion trap experiments \cite{Alonso2016,Johnson2017,McDonnell2007}. The single particle entanglement \cite{Azzini2020} presented here can find applications in the context of
disproving non-contextual hidden variable theories \cite{Hasegawa2011}, and also in the manipulation and interface of quantum the quantum state of an ion. 

% Other proposal exists for the generation of Schrödinger cat states with moderate photon number \cite{Monroe1996}. There are other theoretical proposal to generate cat states \cite{Li1999}, but requiring a measurent of the electronic state.
% Generation of large coherent states is possible \cite{Alonso2016,Johnson2017,McDonnell2007}. 
% Entanglement between motion and spin in ions outside the Lamb-Dicke regime \cite{McDonnell2007}.
% Kerr-type evolution has been proposed in trapped ions \cite{Stobinska2011}.
% Single particle entanglement has been proposed \cite{Azzini2020} and used to demonstrate that quantum mechanics cannot be described by noncontextual hidden variable theories \cite{Hasegawa2011}.
% Other proposals exist for entangling the different degree of freedom of motion in a trapped environment with a quantized electromagnetic mode \cite{Semiao2001}.
% Rabi oscillations might be revived using the echo technique \cite{Meunier2005}. 
\begin{acknowledgments}
This work was supported by CONAHCYT Research Grant CF-2023-I-1751. C.V.-V.  was supported by a postdoctoral grant under the CONAHCYT program {\it Estancias posdoctorales por México 2023(1)}.
\end{acknowledgments}

% Bibliography
%\nocite{*}
\bibliographystyle{apsrev4-1}
\bibliography{references}

\end{document}